\documentclass[twocolumn]{aastex62}

\newcommand{\dsct}{$\delta$~Sct }

\usepackage[]{xcolor}

\received{2019 March 13}
\revised{2019 August 7}
\accepted{2019 August 17}

\submitjournal{ApJL}

\shorttitle{Tidal pulsations in the eclipsing binary U~Gru}

\shortauthors{Bowman et al.}

\begin{document}

\title{DISCOVERY OF TIDALLY-PERTURBED PULSATIONS IN THE ECLIPSING BINARY U~GRU: \\ A PIONEERING SYSTEM FOR TIDAL ASTEROSEISMOLOGY}

\correspondingauthor{Dominic M. Bowman}
\email{dominic.bowman@kuleuven.be}


\author[0000-0001-7402-3852]{Dominic M. Bowman}
\affiliation{Institute of Astronomy, KU Leuven, Celestijnenlaan 200D, 3001 Leuven, Belgium}

\author[0000-0002-3054-4135]{Cole Johnston}
\affiliation{Institute of Astronomy, KU Leuven, Celestijnenlaan 200D, 3001 Leuven, Belgium}

\author[0000-0003-0842-2374]{Andrew Tkachenko}
\affiliation{Institute of Astronomy, KU Leuven, Celestijnenlaan 200D, 3001 Leuven, Belgium}

\author[0000-0001-5094-3910]{David E. Mkrtichian}
\affiliation{National Astronomical Research Institute of Thailand, 260 Moo 4, T. Donkaew, A. Maerim, Chiangmai, 50180 Thailand}

\author[0000-0002-9253-0302]{Khemsinan Gunsriwiwat}
\affiliation{National Astronomical Research Institute of Thailand, 260 Moo 4, T. Donkaew, A. Maerim, Chiangmai, 50180 Thailand}

\author[0000-0003-1822-7126]{Conny Aerts}
\affiliation{Institute of Astronomy, KU Leuven, Celestijnenlaan 200D, 3001 Leuven, Belgium}
\affiliation{Department of Astrophysics, IMAPP, Radboud University Nijmegen, P.O. Box 9010, 6500 GL Nijmegen, the Netherlands}
\affiliation{Max Planck Institute for Astronomy, Koenigstuhl 17, 69117 Heidelberg, Germany}


\begin{abstract}
The interior physics of stars is currently not well constrained for early-type stars. This is particularly pertinent for multiple systems as binary interaction becomes more prevalent for more massive stars, which strongly affects their evolution. High-precision photometry from the {\it Transiting Exoplanet Survey Satellite} (TESS) mission offers the opportunity to remedy the dearth of observations of pulsating stars that show evidence of binary interaction, specifically pulsating mass-accreting components of semi-detached Algol-type eclipsing binary (oEA) systems. We present the TESS light curve of the circular eclipsing binary system U~Gru (TIC~147201138), which shows evidence for free heat-driven pressure modes and a series of tidally-perturbed pressure modes. We highlight the asteroseismic potential of studying pulsating stars in binary systems, and demonstrate how tidal asteroseismology can be applied to infer the influence of binary interaction on stellar structure. 
\end{abstract}

\keywords{asteroseismology -- stars: evolution -- stars: massive -- stars: oscillations -- stars: rotation -- stars: individual (U~Gru)}



\section{Introduction}
\label{section: intro}

For intermediate- and high-mass stars, the incidence of multiplicity and the interaction of stars within binary systems have a substantial impact on their structure and evolution \citep{Sana2012b, Duchene2013b, Moe2017a, DeMarco_O_2017a}. However the effects of binary interaction remain largely unconstrained in stellar models, owing to a lack of systems to characterize and test theoretical predictions. Specifically the consequences of deformation from sphericity, mass transfer, deviation from synchronicity, and tidally induced or perturbed stellar oscillations strongly influence the evolution of stars in multiple systems \citep{Zahn1975, Kumar1995a, Willems_B_2002d, Aerts2004g, Ogilvie_G_2014a, MacLeod2019a}.

By means of asteroseismology -- the study of stellar pulsations -- the interior physics of predominantly single stars covering a range in mass and age in the Hertzsprung--Russell (HR) diagram has been probed \citep{Chaplin2013c, Aerts2015b, Hekker2017a}. Stellar pulsations come in two main flavors based on the dominant restoring force: pressure (p) and gravity (g) modes. The former typically have pulsation periods of order hours and predominantly probe the stellar envelope, whereas the latter have pulsation periods of order days and probe the near-core region \citep{ASTERO_BOOK}. The success of asteroseismology has revealed the internal rotation profiles of tens of intermediate-mass main-sequence stars \citep{Aerts2017b} and thousands of red giant stars \citep{Mosser2012c, Gehan2018}, which demonstrate that a more efficient angular momentum transport mechanism is needed than currently included in evolutionary models of single stars \citep{Aerts2019b*}.

Ground-based photometry is usually sufficient to detect and measure the period of an eclipsing binary system, but asteroseismic studies require continuous, high-precision and long-term photometric time series to extract and identify pulsation modes frequencies \citep{ASTERO_BOOK}. The potential of combining binary and asteroseismic modeling was recently exemplified by \citet{Guo_Z_2016a, Guo_Z_2017a} and \citet{Johnston2019a}, yet few pulsating eclipsing binary systems have suitable space photometry to perform robust modeling. The application of tidal asteroseismology using self-excited free pulsations perturbed by tidal forces or pulsations driven by tidal forces within a binary system offers a unique methodology for directly constraining the impact of tides on stellar structure and evolution. 

More commonly known are tidally-induced pulsations in `heartbeat stars', which are driven by resonances at exact orbital harmonics caused by the temporary tidal distortion at close periastron passage \citep{Welsh2011, Thompson2012}. Approximately 170 of these systems are known \citep{Prsa2011a, Kirk_B_2016}\footnote{binary catalog: \url{http://keplerEBs.villanova.edu}}, and modeling has provided insight of their interior structure \citep{Welsh2011, Beck2014, Smullen2015, Hambleton2013c, Hambleton2018a, Fuller2017c}. An example of such an pulsating eccentric binary system is KIC~4544587, which also contains tidally-split p~modes \citep{Hambleton2013c}.

On the other hand, tidally-perturbed pulsations, which are self-excited free oscillation modes shifted from their eigenfrequencies due to tidal forces, the tidal deformation of the stellar structure, asynchronous rotation or some combination thereof, lack firm observational characterization. Although a theoretical framework has been developed to investigate tidally-perturbed pulsations (see \citealt{Cowling1941, Polfliet1990, Smeyers1998a}), there are few candidate binary systems known to exhibit such variability. More discoveries of eclipsing binary systems with at least one star with tidally-perturbed pulsation modes are clearly of high value to the improvement and development of tidal and stellar evolution theory.

Promising targets for tidal asteroseismology include Algol-type systems in which the mass-accreting primary exhibits p~modes (i.e. oEA systems) having gained a considerable fraction of its mass from an evolved secondary filling its Roche lobe \citep{Mkrtichian2002a, Mkrtichian2004a, Tkachenko2009b, Guo_Z_2016a, Guo_Z_2017a}. The prototype of the oEA class, RZ Cas, revealed strong accretion-driven variability in its pulsation modes \citep{Mkrtichian2018b}, hence motivating renewed observational and theoretical efforts to use pulsation modes to probe the influence of angular momentum transfer and tides on stellar structure and rotation. To date, about 70 oEA systems are known. This number is expected to rapidly increase thanks to the high-precision space photometry from the Transiting Exoplanet Survey Satellite (TESS) mission \citep{Ricker2015}, which is observing a large fraction of the sky with a high photometric precision. In this Letter, we present the TESS light curve for the oEA system U~Gru (TIC~147201138), which has undergone mass transfer and exhibits tidally-perturbed pulsation modes.


\section{The pulsating eclipsing binary U~Gru -- TIC~147201138}
\label{section: U Gru}

In the study of \citet{Brancewicz1980a}, U~Gru was identified as a semi-detached eclipsing binary system with a primary of spectral type A5\,V and an orbital period of 1.88~d. Furthermore, photometry of the system allowed \citet{Brancewicz1980a} to estimate an effective temperature of $T_{\rm eff, 1} \simeq 8000$~K, a mass of $M_{\rm 1} \simeq 2$~M$_{\odot}$ and a radius of $R_{\rm 1} \simeq 2.5$~R$_{\odot}$ for the primary. However, the parameters of the secondary are largely unconstrained.

\begin{figure*}
\includegraphics[width=0.99\textwidth]{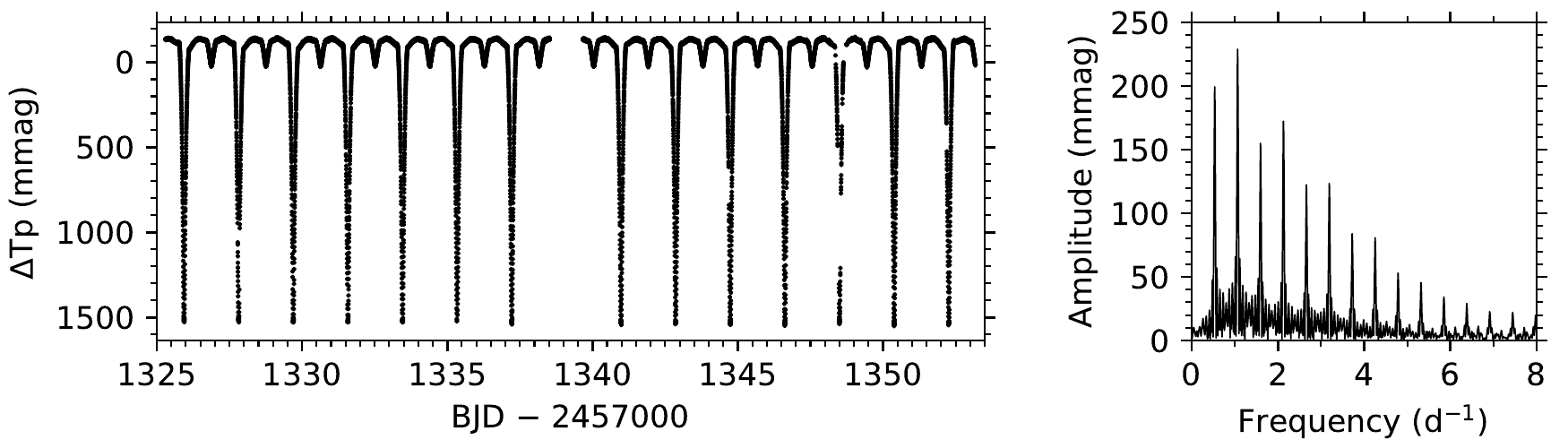}
\caption{{\it Left:} TESS light curve of the pulsating Algol system U~Gru (TIC~147201138). {\it Right:} amplitude spectrum showing the low-frequency orbital harmonic series associated with the eclipses in the light curve.}
\label{figure: LC}
\end{figure*}

The target U~Gru (TIC~147201138) was observed by the TESS mission in its sector~1 and the 2-min cadence light curve spans 27.9~d. We obtained the light curve provided by the TESS Science Team from the Mikulski Archive for Space Telescopes (MAST)\footnote{\url{http://archive.stsci.edu/tess/all_products.html}}. A description of the data processing pipeline of the TESS light curves is provided by \citet{Jenkins2016b}. We extracted the times series in Barycentric Julian Date (BJD -- 2457000), removed obvious outliers, converted the stellar flux into magnitudes, and normalized the light curve to be zero in the mean. The resultant light curve is shown in the left panel of Fig.~\ref{figure: LC}, which shows deep flat-bottomed primary and shallow triangular secondary eclipses. 

We calculated the amplitude spectrum using a Discrete Fourier Transform (DFT; \citealt{Deeming1975, Kurtz1985b}), with a zoom-in of the low-frequency regime shown in the right panel of Fig.~\ref{figure: LC}. The Nyquist frequency of the 2-min TESS light curves is approximately 360~d$^{-1}$ which is far higher than the typical frequency regime of p~modes observed in \dsct stars (i.e. $4 \lesssim \nu \leq 70$~d$^{-1}$; \citealt{Breger2000b, Bowman2018a}), and has the advantage of not introducing significant amplitude suppression of any high-frequency pulsation modes (see \citealt{Bowman_BOOK}). The precise orbital frequency of U~Gru, $\nu_{\rm orb} = 0.531774 \pm 0.000004$~d$^{-1}$ (i.e. $P_{\rm orb} = 1.88050 \pm 0.00001$~d), was determined using a multi-frequency non-linear least-squares fit to the light curve, which included the orbital frequency and the 35 significant consecutive harmonics that have amplitudes larger than $3\sigma$ of the least-squares amplitude error (i.e. $A \geq 119$~$\mu$mag). We determined the linear binary ephemeris of U~Gru using the minimum of the primary eclipse as:
\begin{equation}
{\rm BJD_{min}} = 2458325.93900 + (1.88050 \pm 0.00001)^{\rm d} \times E ~ . 
\label{equation: ephemeris}
\end{equation}

The same orbital frequency was found if 190 harmonics covering up to 100~d$^{-1}$ were included in the multi-frequency non-linear least-squares fit. The significance of the harmonics of the orbital frequency above 20~d$^{-1}$ ranges between $1\sigma$ and $26\sigma$. As expected, the harmonics are exact multiples of the orbital frequency.

\begin{figure*}
\includegraphics[width=0.99\textwidth]{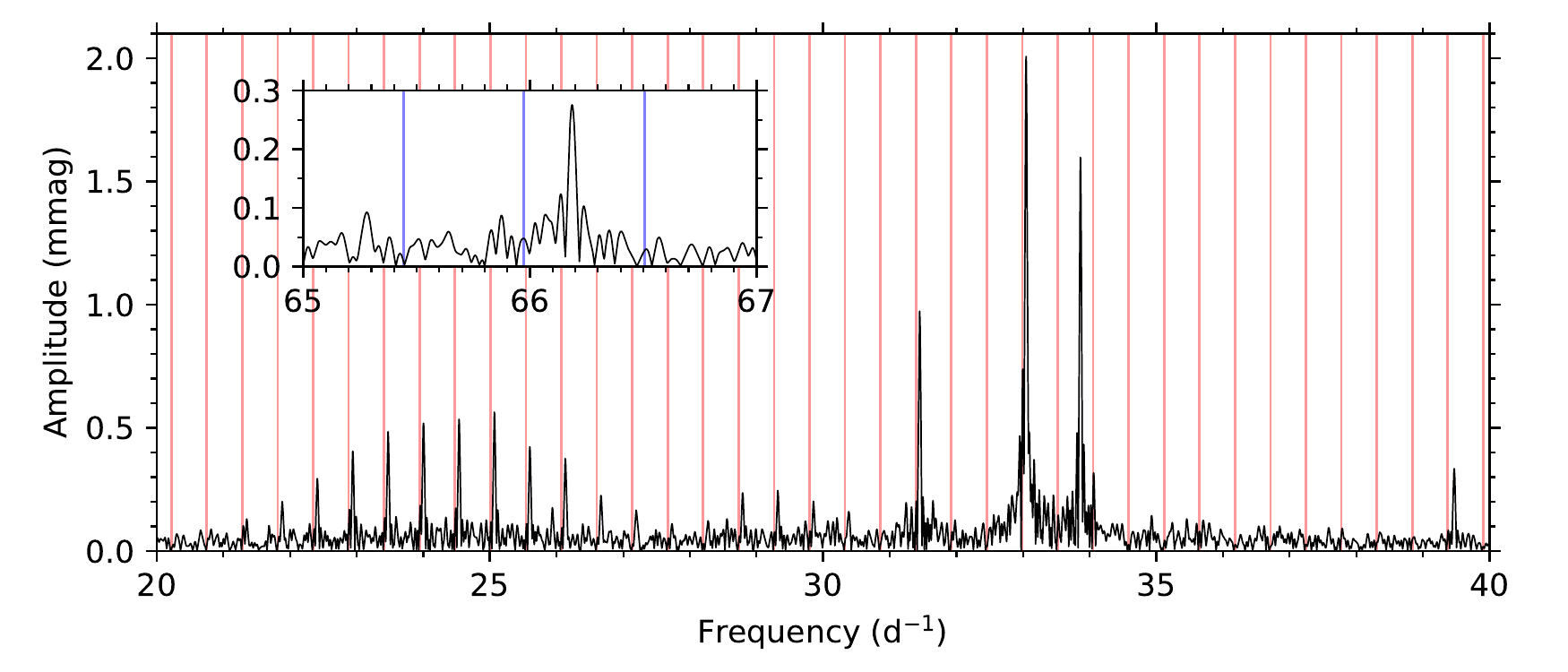}
\caption{Residual amplitude spectrum after pre-whitening the orbital harmonics (denoted as vertical red lines in main panel and blue lines in sub-panel for clarity) in the TESS light curve of U~Gru revealing pulsation mode frequencies.}
\label{figure: FT}
\end{figure*}

The amplitude spectrum of the residual light curve after subtracting the 190-harmonic multi-frequency fit revealed multiple p~mode frequencies, as shown in Fig.~\ref{figure: FT}, in which harmonics of the orbital frequency are indicated by the vertical red lines. The most notable variance in the amplitude spectrum of the residuals is a series of frequencies between 21 and 31~d$^{-1}$, which range in amplitude and are separated by the orbital frequency of the binary system given the frequency uncertainties, yet all are significantly different to integer multiples of the orbital frequency. The harmonics of the orbital frequency in this frequency range are of similar amplitude to the pulsation modes and are significant at $3\sigma$ of the least-squares amplitude error. The extracted pulsation mode frequencies are given in Table~\ref{table: freq list}.

All of the pulsation mode frequencies in the residual amplitude spectrum in Fig.~\ref{figure: FT} are independent as they are resolved from a harmonic of the orbital frequency by more than twice the Rayleigh resolution criterion, which is defined as $1/\Delta\,T = 1/27.88\,{\rm d} = 0.036$~d$^{-1}$. In the sub-panel of Fig.~\ref{figure: FT}, we show a zoom-in of an isolated frequency at $\nu = 66.1853 \pm 0.0023$~d$^{-1}$, which is plotted separately for clarity because of the wide range in pulsation mode frequencies in U~Gru. There are no significant frequencies in the residual amplitude spectrum above 40~d$^{-1}$ except the one at 66.1853~d$^{-1}$.

The pulsation modes in the residual amplitude spectrum of U~Gru were extracted using iterative pre-whitening and optimized using a multi-frequency non-linear least squares fit to the residual light curve \citep{Bowman_BOOK}. In total, 19 significant frequencies in the long series of frequencies were extracted from the residual light curve (cf. Table~\ref{table: freq list}). Although only the first 17 can be considered as part of an unbroken series, the final two frequencies, namely the first and third highest amplitude frequencies in the residual amplitude spectrum, 31.4469 and 33.0442~d$^{-1}$, follow the same pattern of being offset from an orbital harmonic. The frequency offset of a pulsation mode from its adjacent (lower frequency) harmonic of the orbital frequency is also provided in Table~\ref{table: freq list}. The average frequency offset from the adjacent orbital harmonic in the long series of modes is $0.074$~d$^{-1}$, and the deviation from a constant spacing is significant for several frequencies at the 1-$\sigma$ level. For example, the pulsation frequency at 28.2758~d$^{-1}$ has a significantly larger offset of $0.092 \pm 0.006$~d$^{-1}$ from $53\,\nu_{\rm orb}$.

Furthermore, the amplitude spectrum of the residuals shown in Fig.~\ref{figure: FT} contains additional independent pulsation mode frequencies that cannot be associated with the series of frequencies offset from the orbital harmonics, and represent independent p~modes self-excited by the opacity mechanism \citep{Breger2000b, ASTERO_BOOK}. Hence none of the pulsation mode frequencies in U~Gru are at exact integer harmonics of the orbital frequency and many of them appear to be equally spaced by the orbital frequency of the binary system given the frequency uncertainties of the observations.

\begin{deluxetable*}{r r r r r r}
\tablecaption{Frequencies, amplitudes and phases of the significant pulsation modes in U~Gru. $1\sigma$ uncertainties calculated from the multi-frequency non-linear least-squares fit and the signal-to-noise ratio (S/N) of each pulsation mode in the residual amplitude spectrum are given. The measured frequency difference between an independent pulsation mode frequency, $\nu$, and the adjacent (lower frequency) harmonic, $i$, of the orbital frequency, $i\,\nu_{\rm orb}$, is also provided in the last column. \label{table: freq list}}
\tablehead{
\colhead{Frequency} & \colhead{Amplitude} & \colhead{Phase} & \colhead{S/N} & \colhead{$i$} & \colhead{$\nu - i\,\nu_{\rm orb}$} \\
\colhead{(d$^{-1}$)} & \colhead{(mmag)} & \colhead{(rad)} & \colhead{} & \colhead{} & \colhead{(d$^{-1}$)}
}
\startdata
$21.8802 \pm 0.0029$	&	$0.217 \pm 0.032$	&	$0.46 \pm 0.15$	&	$4.27$	&	$41$		&	$0.078 \pm 0.004$	\\
$22.4077 \pm 0.0021$	&	$0.296 \pm 0.032$	&	$-2.88 \pm 0.11$	&	$4.92$	&	$42$		&	$0.073 \pm 0.003$	\\
$22.9411 \pm 0.0015$	&	$0.413 \pm 0.032$	&	$0.15 \pm 0.08$	&	$6.03$	&	$43$		&	$0.075 \pm 0.002$	\\
$23.4696 \pm 0.0013$	&	$0.490 \pm 0.032$	&	$-3.13 \pm 0.07$	&	$6.26$	&	$44$		&	$0.072 \pm 0.002$	\\
$24.0022 \pm 0.0012$	&	$0.520 \pm 0.032$	&	$-0.03 \pm 0.06$	&	$6.37$	&	$45$		&	$0.072 \pm 0.002$	\\
$24.5349 \pm 0.0012$	&	$0.545 \pm 0.032$	&	$2.95 \pm 0.06$	&	$6.37$	&	$46$		&	$0.073 \pm 0.001$	\\
$25.0644 \pm 0.0011$	&	$0.564 \pm 0.032$	&	$-0.41 \pm 0.06$	&	$6.75$	&	$47$		&	$0.071 \pm 0.001$	\\
$25.5979 \pm 0.0015$	&	$0.417 \pm 0.032$	&	$2.59 \pm 0.08$	&	$5.72$	&	$48$		&	$0.073 \pm 0.002$	\\
$26.1303 \pm 0.0016$	&	$0.381 \pm 0.032$	&	$-0.69 \pm 0.08$	&	$5.70$	&	$49$		&	$0.073 \pm 0.002$	\\
$26.6644 \pm 0.0028$	&	$0.225 \pm 0.032$	&	$2.26 \pm 0.14$	&	$4.33$	&	$50$		&	$0.076 \pm 0.003$	\\
$27.1966 \pm 0.0040$	&	$0.155 \pm 0.032$	&	$-1.37 \pm 0.21$	&	$3.51$	&	$51$		&	$0.076 \pm 0.004$	\\
$27.7289 \pm 0.0057$	&	$0.109 \pm 0.032$	&	$0.30 \pm 0.30$	&	$2.72$	&	$52$		&	$0.077 \pm 0.006$	\\
$28.2758 \pm 0.0061$	&	$0.103 \pm 0.032$	&	$-2.95 \pm 0.32$	&	$2.44$	&	$53$		&	$0.092 \pm 0.006$	\\
$28.7910 \pm 0.0025$	&	$0.246 \pm 0.032$	&	$-0.78 \pm 0.13$	&	$3.97$	&	$54$		&	$0.075 \pm 0.003$	\\
$29.3180 \pm 0.0025$	&	$0.247 \pm 0.032$	&	$2.16 \pm 0.13$	&	$4.12$	&	$55$		&	$0.071 \pm 0.003$	\\
$29.8512 \pm 0.0035$	&	$0.181 \pm 0.032$	&	$-1.34 \pm 0.18$	&	$3.12$	&	$56$		&	$0.072 \pm 0.004$	\\
$30.3800 \pm 0.0037$	&	$0.167 \pm 0.032$	&	$1.69 \pm 0.19$	&	$2.77$	&	$57$		&	$0.069 \pm 0.004$	\\
$31.4469 \pm 0.0006$	&	$0.984 \pm 0.032$	&	$1.01 \pm 0.03$	&	$8.17$	&	$59$		&	$0.072 \pm 0.001$	\\
$33.0442 \pm 0.0003$	&	$1.997 \pm 0.032$	&	$-2.16 \pm 0.02$	&	$8.22$	&	$62$		&	$0.074 \pm 0.001$	\\
\hline
$33.8598 \pm 0.0004$	&	$1.571 \pm 0.032$	&	$0.95 \pm 0.02$	&	$8.78$	&	$63$		&	$0.358 \pm 0.001$	\\
$39.4689 \pm 0.0020$	&	$0.323 \pm 0.032$	&	$2.24 \pm 0.10$	&	$6.72$	&	$74$		&	$0.112 \pm 0.002$	\\
$66.1853 \pm 0.0023$	&	$0.276 \pm 0.032$	&	$0.41 \pm 0.12$	&	$5.97$	&	$124$	&	$0.245 \pm 0.009$	\\
\enddata
\end{deluxetable*}


\section{Discussion}
\label{section: discussion}

\begin{figure*}
\includegraphics[width=0.99\textwidth]{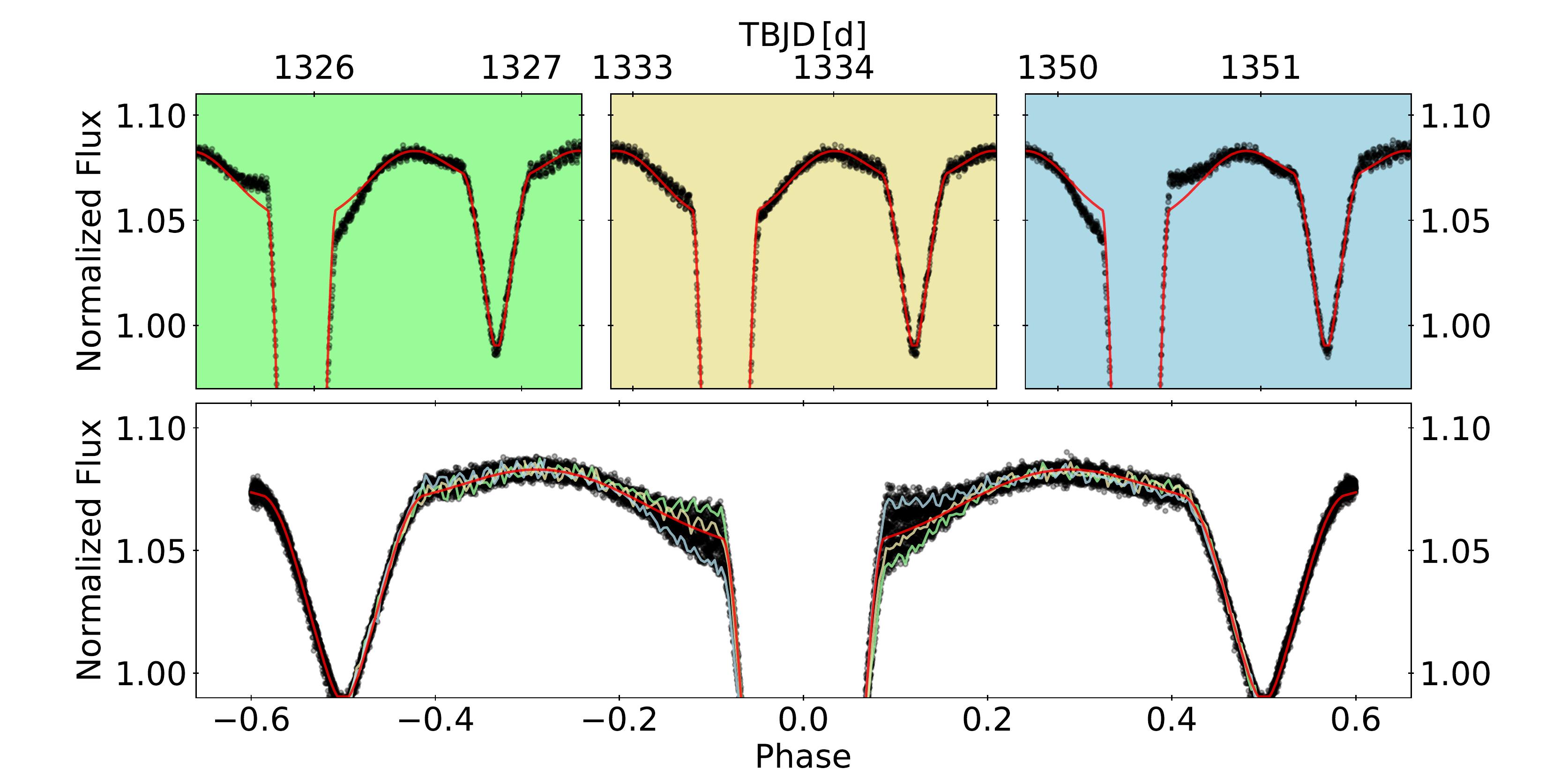}
\caption{{\it Bottom panel:} binary model (red line) and the phase-folded TESS observations of U~Gru (black points). {\it Top panels:} binary model for three sections of the TESS data demonstrating how the flux and shape of the light curve at ingress and egress in the primary eclipse changes throughout the observations. The flux offset is largest at the start (green panel) and end (blue panel) of the TESS light curve and reverses symmetry in the middle (yellow panel). The light curves from the three top panels are overplotted in the bottom panel to emphasize the changing flux at ingress and egress of the primary eclipse.}
\label{figure: binary model}
\end{figure*}

We analysed the TESS light curve of U~Gru with the \texttt{PHOEBE} eclipsing binary modeling code\footnote{\url{http://phoebe-project.org/}} \citep{Prsa2005c} using a semi-detached configuration in which the secondary fills its Roche lobe and an assumption of zero eccentricity. The phase-folded TESS data and the binary model are shown as the black points and red line, respectively, in the bottom panel of Fig.~\ref{figure: binary model}. Binary modeling and the totality of the primary eclipses confirms a near edge-on inclination, and indicates a probable photometric mass ratio of $q \simeq 0.17$, although the latter is unconstrained without spectroscopy. These results support the literature classification of U~Gru being an Algol-like system \citep{Brancewicz1980a} and validate the discovery that U~Gru is a new member of the oEA class of variable stars.

The long series of pulsation modes in U~Gru spans more than 17 frequencies, and the offset of these modes from orbital harmonics precludes a scenario in which they are excited in a `heartbeat' scenario. The frequency spacing within the long series of pulsation modes cannot be identified as a large frequency separation -- i.e. the difference in frequency of consecutive high radial order modes of the same angular degree -- using stellar structure models of single stars and the estimated parameters of both components \citep{Brancewicz1980a, ASTERO_BOOK}. However, owing to the mass transfer that has occurred and the need for including the effect of tides to stellar structure, estimates of the expected large and small frequency separations are uncertain for such a system. Nonetheless, the discovery of a long series of pulsation modes separated by the orbital frequency of the binary system demonstrates the importance of tides for the component stars of U~Gru.

Perhaps the most interesting aspect of U~Gru, is the changing shape of the light curve at the phase of ingress and egress of the primary eclipse in comparison to the best fitting binary model, which is demonstrated in Fig.~\ref{figure: binary model}. In the phase-folded light curve of U~Gru, it appears as if the pulsation amplitudes are growing near primary eclipse, but as demonstrated in Fig.~\ref{figure: binary model}, this is caused by the over-plotting of the changing normalized flux at ingress and egress at primary eclipse. Such an effect could be explained by the change in flux caused by a co-rotating surface feature on a component star within an asynchronous system, but testing this hypothesis must await phase-resolved high-resolution spectroscopy.

We also note that the highest amplitude pulsation mode in the residual amplitude spectrum at 33.0442~d$^{-1}$ is exhibiting amplitude modulation (see e.g. \citealt{Bowman2016a}) on a time scale longer than the orbital period. Amplitude modulation of pulsation modes both during the orbital phase and on time scales of hundreds to thousands of days has been inferred in some oEA stars using ground-based data (e.g. RZ~Cas; \citealt{Mkrtichian2018b}). Accretion-driven variability and variable extinction within the semi-transparent circumbinary environment is plausibly the cause of long-term amplitude variability in U~Gru, as has been inferred for other oEA systems (e.g. \citealt{Mkrtichian2018b}).

\begin{figure*}
\includegraphics[width=0.99\textwidth]{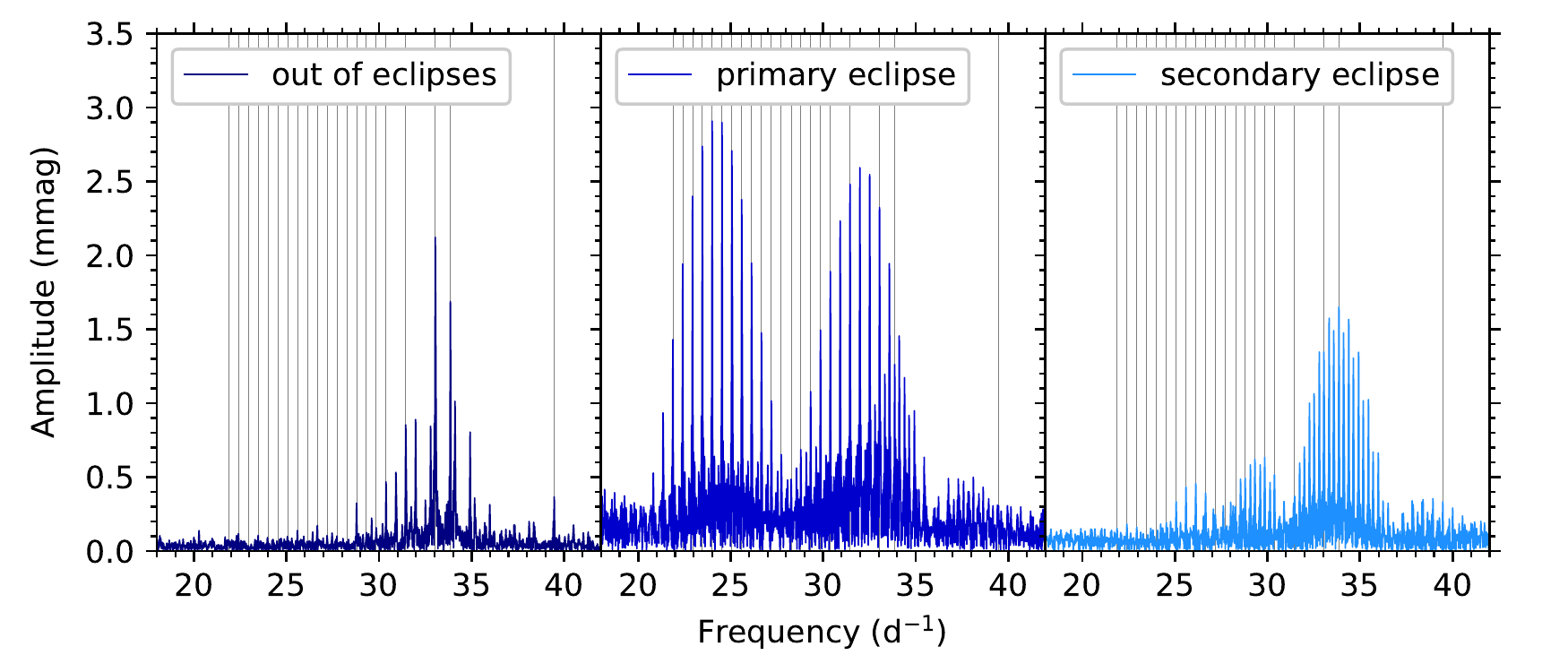}
\caption{Amplitude spectra of the residual TESS light curve (i.e. after subtracting the multi-harmonic binary model) of U~Gru using data from different binary phases (in primary: $-0.09 \leq \varphi \leq 0.09$ and in secondary: $0.41 \leq \varphi \leq 0.59$; cf. Fig.~\ref{figure: binary model}). The vertical grey lines indicate the frequencies included in Table~\ref{table: freq list} identified as significant from the entire light curve.}
\label{figure: AMod}
\end{figure*}

Furthermore, the amplitudes of the pulsation modes comprising the long series of modes in U~Gru are also modulated during the binary orbit, as demonstrated in Fig.~\ref{figure: AMod}, yet the high-frequency independent pulsation modes at 33.8598~d$^{-1}$ and 39.4689~d$^{-1}$ remain approximately constant in amplitude. This demonstrates that the amplitude of the pulsation modes in the long series of modes depends on the binary phase and can be explained by the influence of tides and/or the observed pulsation mode geometry. The amplitudes of the long series of modes are maximal and form multiplet structures in the amplitude spectrum during primary eclipse. Therefore, the series of equally-spaced pulsation modes may originate in the cooler secondary such that the observed modulation is being caused by geometric cancellation and/or the dependence of pulsation mode amplitudes on the changing flux ratio of the two components at a specific binary phase.

Synchronicity and circularisation are typically assumed for post-mass transfer systems (see \citealt{Ogilvie_G_2014a, Guo_Z_2016a, Guo_Z_2017a}), but this is not always the case \citep{Escorza2019b}. The TESS observations of U~Gru indicate an eccentricity of nearly, if not exactly, zero yet the influence of tides are clearly important in U~Gru. \citet{MacLeod2019a} recently investigated the scenario of tides in asynchronous binaries at the Roche limit using hydrodynamical simulations. The resultant tidal forcing frequency, which is determined by the difference between the orbital and envelope rotation frequencies, may coincide with eigenmode frequencies as the orbit shrinks. It was found that waves of decreasing angular degrees ($\ell$) and azimuthal orders ($m$) are resonantly excited as the system evolves within an unstable mass-transfer scenario \citep{MacLeod2019a}. However, such an evolution-based phenomenon is unlikely to be detected in a short light curve spanning only 28~d.

Given these TESS observations, we discuss three mechanisms that could be responsible for the rich pulsation spectrum, which require tidal deformation of the pulsation cavity within a star given the circular orbit of U Gru:

{\it (i) tidally-excited modes:} the tide generating potential of the U~Gru system excites a long series of consecutive radial order modes, in which the particular range in radial order ($n$) provides information on the distorted pulsation cavity \citep{Ogilvie_G_2014a}. However, the long series of modes are significantly offset from harmonics of the orbital frequency making this scenario unfavorable.

{\it (ii) pulsation mode geometry:} the effect of tides act as a perturbation to the pulsation modes which causes amplitude modulation with respect to the observer, resulting in a multiplet split by the orbital frequency. A tidal scenario in which a pulsation mode is trapped to a single hemisphere of the primary and results in amplitude modulation during the orbit and an equally-split multiplet in the amplitude spectrum is plausible for U~Gru, but such a mechanism does not explain the additional independent pulsation modes. On the other hand, the long series of pulsation modes spaced by the orbital frequency may originate from the secondary, whose pulsation mode amplitudes are maximal during primary eclipse because of geometric cancellation and/or the dependence of pulsation mode amplitudes on the changing relative flux ratio at a given binary phase. If this is the case, U~Gru represents a circular eclipsing binary system which contains two pulsating components whose pulsations are influenced by tides. 

{\it (iii) tidally-perturbed modes:} the observed frequencies are free oscillations of consecutive radial order that are self-excited by the opacity mechanism, but whose eigenfrequencies are perturbed from the corresponding single-star unperturbed eigenfrequencies because of the tidal deformation of the pulsation cavity \citep{Polfliet1990, Reyniers_PHD, Reyniers2003a, Reyniers2003b}.

The detection of a long series of pulsation modes, which are significantly different to harmonics of the orbital frequency, in addition to independent free self-excited p~modes indicates that cases (ii) and (iii) are important in U Gru. Therefore the discovery of such a rich pulsation spectrum in an oEA system such as U~Gru offers a unique and exciting prospect for the application of tidal asteroseismology, as the influence of tides can be directly measured from the perturbations to observed pulsation mode frequencies. 

The case for asynchronous rotation being the cause of the tidal torque in a circular binary system will be tested using a spectroscopic measurement of rotation, i.e. $v\,\sin\,i$, and the subsequent investigation of the synchronicity parameter in future binary modeling including spectroscopic radial velocities. We are currently gathering the necessary high-resolution spectroscopy to determine accurate stellar parameters for both components and perform robust binary modeling by combining radial velocity measurements with the TESS light curve of U~Gru. The resultant binary parameters and subsequent asteroseismic modeling will ascertain the most likely overall tidal scenario (Johnston et al., {\it in prep}). The combination of high-precision TESS photometry, high-resolution spectroscopy and forward seismic modeling will provide quantitative measures of the influence of tides on stellar structure and evolution in a pulsating eclipsing binary system.


\acknowledgments
{\it Acknowledgments:} The authors thank the referee, Jim Fuller, for his comments which improved the presentation and interpretation in this work. The TESS data presented in this paper were obtained from the Mikulski Archive for Space Telescopes (MAST) at the Space Telescope Science Institute (STScI), which can be accessed via \dataset[https://doi.org/10.17909/t9-dcby-xt10]{https://doi.org/10.17909/t9-dcby-xt10}. STScI is operated by the Association of Universities for Research in Astronomy, Inc., under NASA contract NAS5--26555. Support to MAST for these data is provided by the NASA Office of Space Science via grant NAG5--7584 and by other grants and contracts. Funding for the TESS mission is provided by the NASA Explorer Program. This research has made use of the SIMBAD database, operated at CDS, Strasbourg, France; the SAO/NASA Astrophysics Data System; and the VizieR catalogue access tool, CDS, Strasbourg, France. The research leading to these results has received funding from the European Research Council (ERC) under the European Union’s Horizon 2020 research and innovation programme (grant agreement N$^\circ$670519: MAMSIE), from the KU\,Leuven Research Council (grant C16/18/005: PARADISE), from the Research Foundation Flanders (FWO) under grant agreement G0H5416N (ERC Runner Up Project), from the BELgian federal Science Policy Office (BELSPO) through PRODEX grant PLATO, as well as from the Research Foundation Flanders (FWO) under grant agreement G0A2917N (BlackGEM). DEM is supported by the National Astronomical Research Institute of Thailand (NARIT), Ministry of Science and Technology of Thailand.


\bibliography{/Users/Dominic/Documents/RESEARCH/Bibliography/master_bib.bib}


\end{document}